\documentclass[aps,prl,twocolumn,preprintnumbers,superscriptaddress,showpacs,floatfix, nofootinbib]{revtex4-1}

\usepackage[caption=false]{subfig}
\usepackage{hyperref}
\usepackage{color}
\usepackage{xcolor}
\usepackage{graphicx}
\usepackage{amsmath}	
\usepackage{siunitx}
\graphicspath{
  {./},
}

\usepackage{amstext}
\usepackage{array}

\usepackage[normalem]{ulem}

\newcommand{\Z}{\rm Z}
\newcommand{\pTZ}{p_T^{\rm Z}}
\newcommand{\pTin}{p_T^{\rm in}}
\newcommand{\pTMC}{p_T^{\rm MC}}
\newcommand{\pTHI}{p_T^{\rm AA}}
\newcommand{\pTvac}{p_T^{\rm pp}}
\newcommand{\pTq}{p_T^{\rm quant}}

\newcommand{\Eq}[1]{Eq.~\eqref{#1}}
\newcommand{\Fig}[1]{Fig.~\ref{#1}}
\newcommand{\Ref}[1]{Ref.~\cite{#1}}
\newcommand{\df}{\text{d}}

\begin{document} 

\title{Sorting out quenched jets}

\author{Jasmine Brewer}
\email{jtbrewer@mit.edu}
\affiliation{Center for Theoretical Physics, Massachusetts Institute of Technology, Cambridge, MA 02139, USA}

\author{Jos\'e Guilherme Milhano}
\email{guilherme.milhano@tecnico.ulisboa.pt}
\affiliation{LIP, Av.\ Prof.\ Gama Pinto, 2, P-1649-003 Lisboa, Portugal}
\affiliation{Instituto Superior T\'ecnico (IST), Universidade de Lisboa, Av.\ Rovisco Pais 1, 1049-001, Lisbon, Portugal}

\author{Jesse Thaler}
\email{jthaler@mit.edu}
\affiliation{Center for Theoretical Physics, Massachusetts Institute of Technology, Cambridge, MA 02139, USA}
\affiliation{Department of Physics, Harvard University, 17 Oxford Street, Cambridge, MA 02138, USA}

\begin{abstract}
We introduce a new ``quantile'' analysis strategy to study the modification of jets as they traverse through a droplet of quark-gluon plasma.
To date, most jet modification studies have been based on comparing the jet properties measured in heavy-ion collisions to a proton-proton baseline at the same reconstructed jet transverse momentum ($p_T$).
It is well known, however, that the quenching of jets from their interaction with the medium leads to a migration of jets from higher to lower $p_T$, making it challenging to directly infer the degree and mechanism of jet energy loss.
Our proposed quantile matching procedure is inspired by (but not reliant on) the approximate monotonicity of energy loss in the jet $p_T$.
In this strategy, jets in heavy-ion collisions ordered by $p_T$ are viewed as modified versions of the same number of highest-energy jets in proton-proton collisions, and the fractional energy loss as a function of jet $p_T$ is a natural observable ($Q_{\rm AA}$).
Furthermore, despite non-monotonic fluctuations in the energy loss, we use an event generator to validate the strong correlation between the $p_T$ of the parton that initiates a heavy-ion jet and the $p_T$ of the vacuum jet which corresponds to it via the quantile procedure ($\pTq$).
We demonstrate that this strategy both provides a complementary way to study jet modification and mitigates the effect of $p_T$ migration in heavy-ion collisions.

\end{abstract}

\preprint{MIT-CTP/5089}
\maketitle

The deconfined phase of QCD matter, the quark-gluon plasma, was first discovered in collisions of heavy nuclei at the Relativistic Heavy Ion Collider \cite{Adler:2001nb,Arsene:2004fa,Back:2004je,Adcox:2004mh,Adams:2005dq} and confirmed at the Large Hadron Collider \cite{Aamodt:2010pa,Aad:2010bu,Chatrchyan:2011sx}.
As in high-energy proton-proton collisions, heavy-ion collisions produce collimated sprays of particles, called jets, from highly energetic scatterings of quarks and gluons.
The observation of ``jet quenching''---a strong suppression and modification of jets in heavy-ion collisions \cite{Aad:2010bu,Chatrchyan:2011sx,Adam:2015ewa}---ushered in a new era of studying the properties of the quark-gluon plasma by measuring its effect on jets \cite{Appel:1985dq,Blaizot:1986ma,Gyulassy:1990ye,Wang:1991xy,Chatrchyan:2012nia,Chatrchyan:2012gt,Chatrchyan:2013exa,Chatrchyan:2014ava,Khachatryan:2015lha,Aaboud:2017eww,Acharya:2017goa,Aaboud:2018twu,Aaboud:2018hpb,Acharya:2018uvf}.

A central issue in interpreting jet quenching measurements is that medium-induced modifications necessarily affect how jets are identified experimentally.
Current methods compare proton-proton and heavy-ion jets of the same final (reconstructed) transverse momentum $p_T$ and, as such, inevitably suffer from significant biases from the migration of jets from higher to lower $p_T$ due to medium-induced energy loss (see \cite{CasalderreySolana:2007pr,dEnterria:2009xfs}).
While these methods have been very successful in qualitatively demonstrating the phenomena of jet quenching, quantitive studies often necessitate interpreting the data through theoretical models which include migration effects.
Ideally, one would like to isolate samples of jets in proton-proton and heavy-ion collisions which were statistically equivalent when they were produced, differing only by the effects of the plasma.

\begin{figure*}
\subfloat[\label{fig:quant_a}]{%
  \includegraphics[width=.5\linewidth]{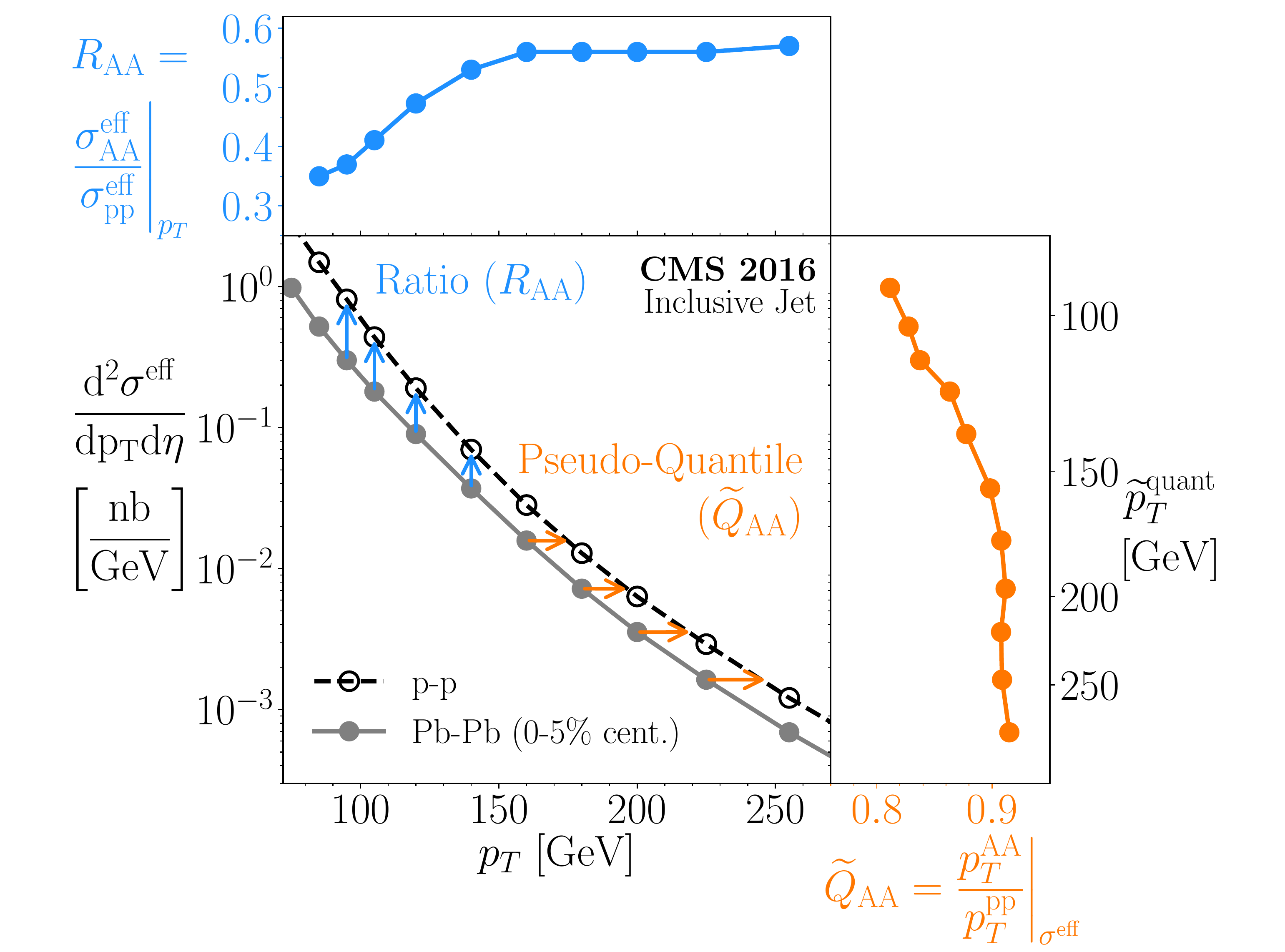}%
}
\subfloat[\label{fig:quant_b}]{%
  \includegraphics[width=.5\linewidth]{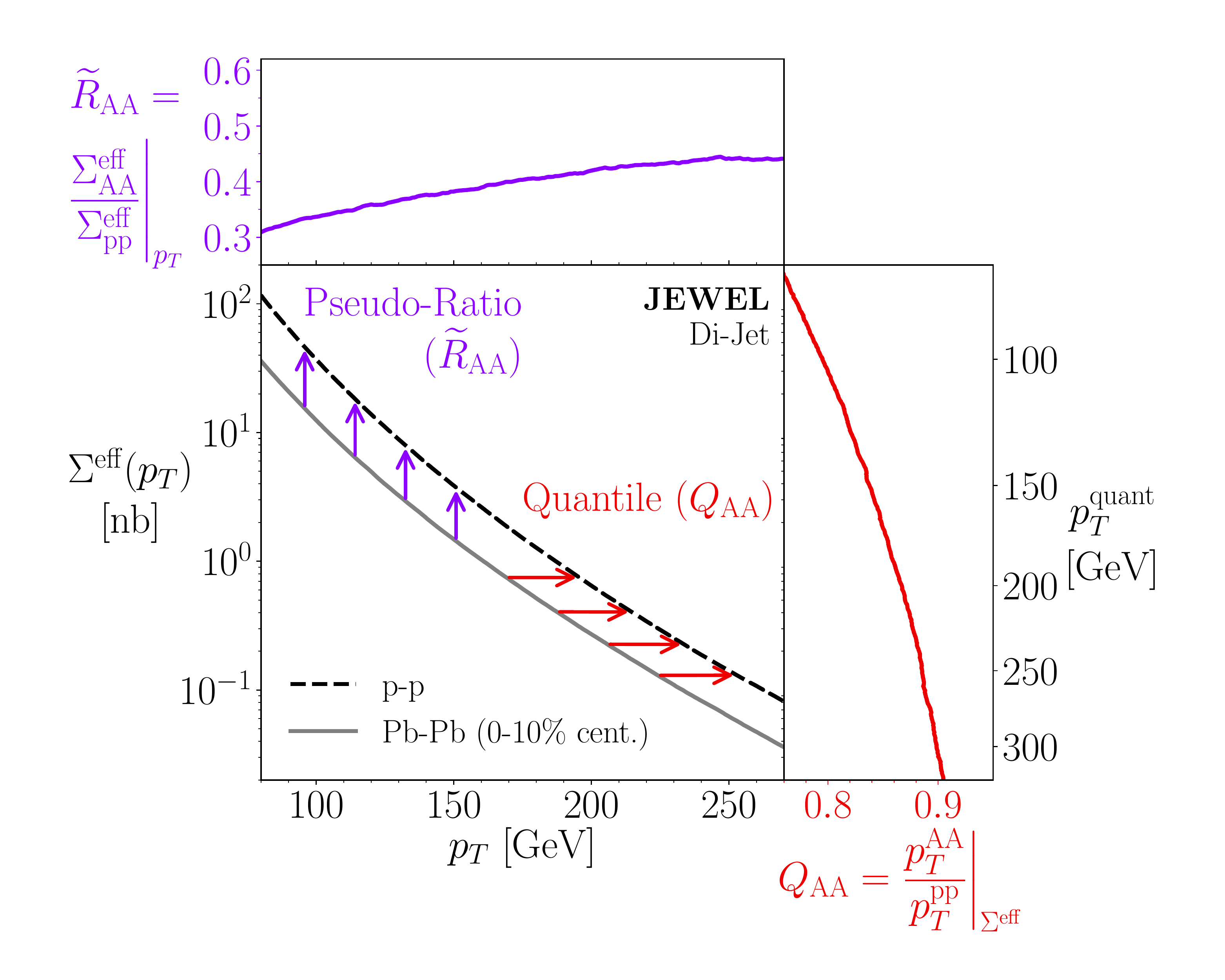}%
}
\caption{\label{fig:quant}
Illustration comparing the ratio and quantile procedures.
(a)
The inclusive jet $p_T$ spectra measured by CMS \cite{Khachatryan:2016jfl}, for a jet radius of $R=0.4$.
The standard jet ratio $R_{\rm AA}$ (blue) compares heavy-ion and proton-proton jet cross-sections vertically at the same reconstructed jet $p_T$.
(b)
The jet $p_T$ cumulative cross-sections extracted from \textsc{Jewel} \cite{Zapp:2013vla,KunnawalkamElayavalli:2016ttl}.
The quantile procedure $Q_{\rm AA}$ (red) compares heavy-ion and proton-proton jet $p_T$ thresholds horizontally at the same cumulative cross-section.
From this, one can map each $\pTHI$ (base of red arrows) into the $p_T$ of proton-proton jets in the same quantile, $\pTq$ (tip of red arrows).
For completeness, we also show the pseudo-quantile $\widetilde{Q}_{\rm AA}$ (orange, with corresponding $\widetilde{p}_T^{\rm quant}$) defined on the cross-section and pseudo-ratio $\widetilde{R}_{\rm AA}$ (purple) defined on the cumulative cross-section.
Though we will not explore the use of $\widetilde{Q}_{\rm AA}$ or $\widetilde{R}_{\rm AA}$ in the present study, we note that in \textsc{Jewel}, the values of $\pTq$ and $\widetilde{p}_T^{\rm quant}$ differ by only a few percent.
}
\label{fig:quantile-cartoon}
\end{figure*}

In this Letter, we propose a novel data-driven strategy for comparing heavy-ion ($\rm AA$) jet measurements to proton-proton ($\rm pp$) baselines which mitigates, to a large extent, the effect of $p_T$ migration.
The famous jet ratio $R_{\rm AA}$ compares the effective cross-section for jets in proton-proton and heavy-ion collisions with the same reconstructed $p_T$:
\begin{equation}
	R_{\rm AA} = \left. \frac{\sigma^{\rm eff}_{\rm AA}}{\sigma^{\rm eff}_{\rm pp}} \right|_{p_T},
\end{equation} 
as illustrated in blue in \Fig{fig:quant_a}.
Here, we introduce a ``quantile'' procedure, which divides jet samples sorted by $p_T$ into quantiles of equal probability.
Our new proposed observable for heavy-ion collisions is the $p_T$ ratio between heavy-ion and proton-proton jets in the same quantile:
\begin{equation}
\label{eq:QAAdef}
	Q_{\rm AA} = \left. \frac{\pTHI}{\pTvac} \right|_{\Sigma^{\rm eff}},
\end{equation} 
as illustrated in red in \Fig{fig:quant_b}, where $1-Q_{\rm AA}$ is a proxy for the average fractional jet energy loss.
($Q_{\rm AA}$ is not related to $Q_{\rm pA}$ used by ALICE \cite{Toia:2014wia}).
Although $R_{\rm AA}$ can be obtained from $Q_{\rm AA}$ if the proton-proton jet spectrum is known, we will see that the physics interpretation of $R_{\rm AA}$ and $Q_{\rm AA}$ can be quite different.
\Fig{fig:quant_a} additionally shows the pseudo-quantile $\widetilde{Q}_{\rm AA}$, which is related to the observable $S_{\rm loss}$ introduced by PHENIX for single hadrons~\cite{Adler:2006bw,Adare:2012wg,Adare:2015cua}.

To give an intuitive understanding of \Eq{eq:QAAdef}, consider a simplified scenario where medium-induced energy loss is monotonic in the $p_T$ of the initial unquenched jet.
In that case, the $n^{\rm th}$ highest energy jet in a heavy-ion sample is a modified version of the $n^{\rm th}$ highest energy jet in the corresponding proton-proton sample.
Thus, in this simplified picture of energy loss, we can obtain a sample of heavy-ion jets that is statistically equivalent to its proton-proton counterpart by selecting jets with the same (upper) cumulative effective cross-section:
\begin{equation}
	\label{eq:cumXSdef}
\Sigma^{\rm eff}(p^{\rm min}_T) = \int_{p^{\rm min}_T}^{\infty} \df p_T \, \frac{\df \sigma^{\rm eff}}{\df p_T}.
\end{equation}
Note that for comparison to proton-proton cross-sections, heavy-ion cross-sections must be rescaled by the average number of nucleon-nucleon collisions $\langle N_{\rm coll} \rangle$: $\sigma^{\rm eff}_{\rm pp} = \sigma_{\rm pp}$, $\sigma^{\rm eff}_{\rm AA} = \sigma_{\rm AA}/\langle N_{\rm coll} \rangle$.
Of course, energy loss is not strictly monotonic in $p_T$, since other properties of a jet and of the jet-medium interaction influence its energy loss and cause jets with the same initial $p_T$ to lose different fractions of their energy.
Below, we will quantify the usefulness of this quantile picture in the context of a realistic event generator where significant non-monotonicities are indeed present.

Due to the steeply-falling jet production spectrum ($\sigma \sim p_T^{-6}$), jets within a given range in reconstructed heavy-ion $p_T$ are dominated by those which were least modified (see e.g.~\cite{Andrews:2018jcm}).
Addressing this issue requires comparing jets that had the same $p_T$ when they were initially produced.
In rarer events where an energetic $\gamma$ or $\Z$ boson is produced back-to-back with a jet, the unmodified boson energy approximates the initial energy of the recoiling jet \cite{Chatrchyan:2012gt,Sirunyan:2017jic}.
In general jet events, however, the jet energy before medium effects cannot be measured.

A key result of this work is that the quantile picture also provides a natural proxy for the unmodified jet $p_T$ that is observable in general jet events.
Given a heavy-ion jet with reconstructed momentum $\pTHI$, we can define $\pTq$ implicitly as the momentum of a proton-proton jet with the same (upper) cumulative cross-section:
\begin{equation}
	\label{eq:pTquant}
	\Sigma^{\rm eff}_{\rm pp}(\pTq) \equiv \Sigma^{\rm eff}_{\rm AA}(\pTHI).
\end{equation}
In this quantile picture, $\pTq$ is viewed as the initial jet $p_T$ prior to medium effects.
The mapping from $\pTHI$ to $\pTq$ is illustrated by the red arrows in \Fig{fig:quant_b}, with $\pTHI = \pTq \, Q_{\rm AA}(\pTq)$.
Intriguingly, we will show that $\pTq$ approximates the $p_T$ of a heavy-ion jet before quenching with comparable fidelity to the unmodified boson energy $\pTZ$ available only in rarer $\Z$+jet events. 
In particular, comparing properties of proton-proton and heavy-ion jet samples with the same $\pTq$ may substantially enhance the sensitivity of modification observables by targeting jets that were more strongly modified.

\begin{figure*}
\subfloat[\label{fig:pTloss_a}]{%
  \includegraphics[width=.5\linewidth]{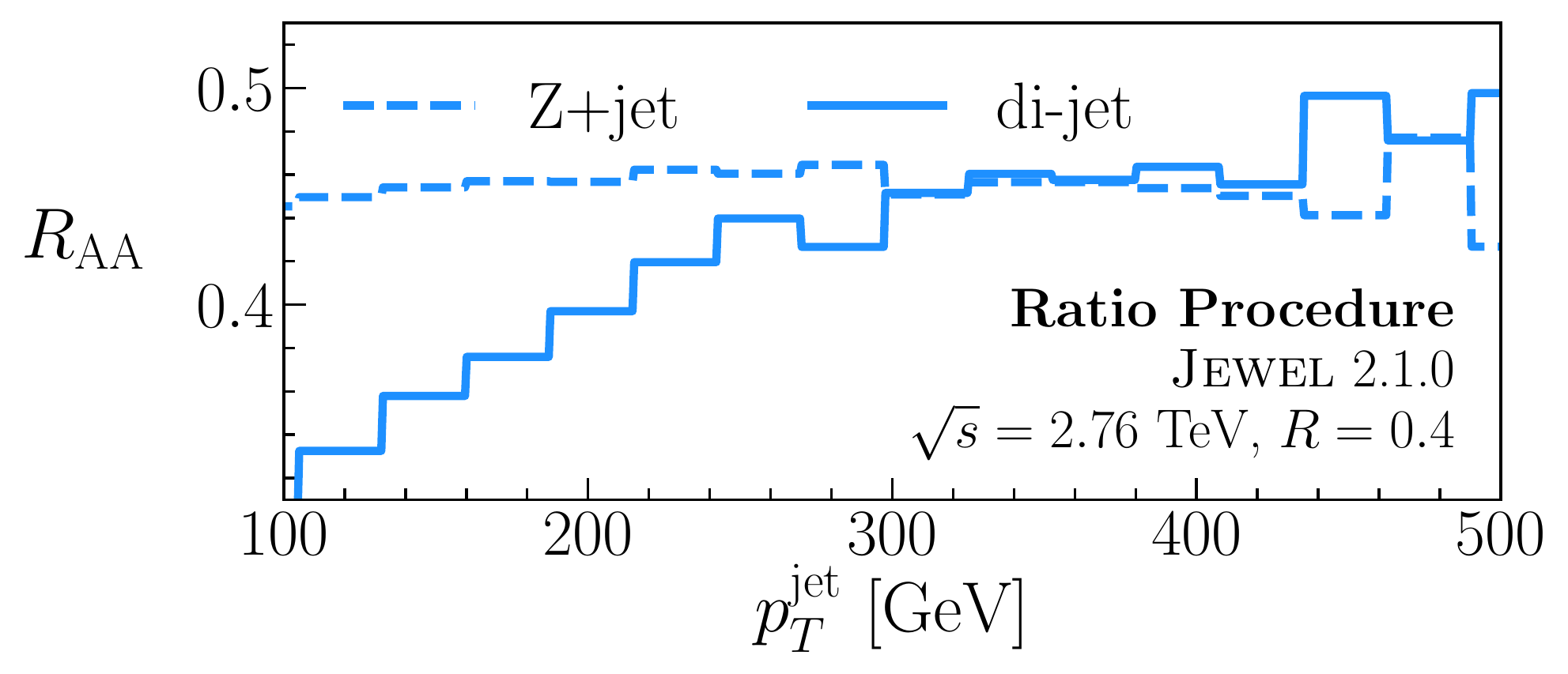}%
}
\subfloat[\label{fig:pTloss_b}]{%
  \includegraphics[width=.5\linewidth]{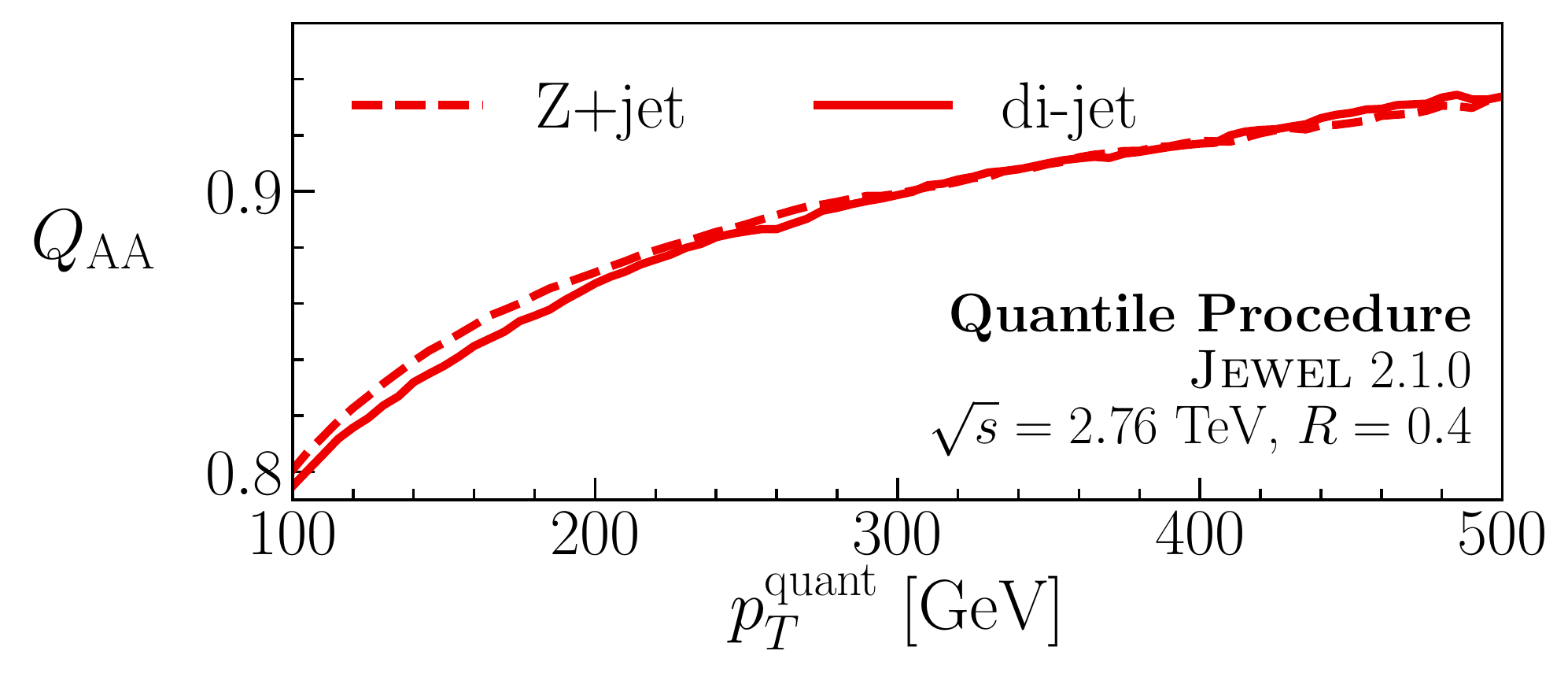}%
}         
\caption{Distributions of (a) $R_{\rm AA}$ as a function of $p_T^{\rm jet}$ and (b) $Q_{\rm AA}$ as a function of $\pTq$, for the $\Z$+jet (dashed) and di-jet (solid) samples in \textsc{Jewel}.
Although $R_{\rm AA}$ and $Q_{\rm AA}$ are derived from the same underlying jet $p_T$ spectra, they provide different and complementary information.
For example, the $p_T$ dependence of $R_{\rm AA}$ is very different for $\Z$+jet and di-jet events in \textsc{Jewel}, while the average fractional $p_T$ loss $1-Q_{\rm AA}$ is similar.
Note that $R_{\rm AA}$ requires binning of the data, while $Q_{\rm AA}$, which is based on the cumulative cross-section, can be plotted unbinned.
}
\label{fig:pTloss}
\end{figure*}

For the remainder of this work, we consider samples of $\Z$+jet and di-jet events in the heavy-ion Monte Carlo event generator \textsc{Jewel} 2.1.0 \cite{Zapp:2013vla,KunnawalkamElayavalli:2016ttl}, based on vacuum jet production in \textsc{Pythia} 6~\cite{Sjostrand:2006za}.
For each process, we generate $2$ million each of proton-proton and head-on ($0-10\%$ centrality) heavy-ion events at $\SI{2.76}{TeV}$ and reconstruct anti-$k_t$ jets using \textsc{FastJet} 3.3.0 \cite{Cacciari:2008gp,Cacciari:2011ma} with radius parameter $R=0.4$ and pseudorapidity $|\eta|<2$.
We include initial state radiation but do not include medium recoils, since medium response is not expected to have a significant effect on \Eq{eq:cumXSdef} at the values of $p_T^{\rm min}$ considered here.
For $\Z$+jet events we identify the $\Z$ from its decay to muons and consider the leading recoiling jet, and for di-jet events we consider the two highest-$p_T$ jets.
We consider $\Z$+jet instead of $\gamma$+jet events to avoid introducing additional cuts to isolate prompt photons which could bias the validation.
The default heavy-ion background in \textsc{Jewel} is a Bjorken expanding medium with initial peak temperature $T_i = \SI{485}{MeV}$ and formation time $\tau_i=\SI{0.6}{fm}$, consistent with the parameters used to fit data at $\SI{2.76}{TeV}$ in more realistic hydrodynamic simulations \cite{KunnawalkamElayavalli:2016ttl,Shen:2012vn}.

Using these $\Z$+jet and di-jet samples from \textsc{Jewel}, \Fig{fig:pTloss_a} shows the standard $R_{\rm AA}$ (also called $I_{\rm AA}$ for $\Z$+jet) and \Fig{fig:pTloss_b} shows the $p_T$ ratio $Q_{\rm AA}$.
Although the $R_{\rm AA}$ for $\Z$+jet and di-jet events have significantly different $p_T$-dependence, it is interesting that the average fractional energy loss of jets is very similar, as quantified by $1 - Q_{\rm AA}$.
This might be surprising since $\Z$+jet and di-jet events have different fractions of quark and gluon jets, though \Ref{Apolinario:2018rhj} suggests that quark and gluon jets may experience similar energy loss in \textsc{Jewel}; whether this is borne out in data is an open question.
Regardless, it is clear that $R_{\rm AA}$ and $Q_{\rm AA}$ offer complementary probes of the jet quenching phenomenon and are therefore both interesting observables in their own right. 
The quantile procedure also shows that the highest-$p_T$ jets lose a small fraction of their energy on average ($(1 - Q_{\rm AA}) \sim 5\%$), even though $R_{\rm AA}$ is far below one.
This result can be compared to other methods for extracting the average energy loss from data, for example \Ref{He:2018gks}.

\begin{figure*}
\subfloat[\label{fig:pT-vs-pTin_a}]{%
  \includegraphics[width=.5\linewidth]{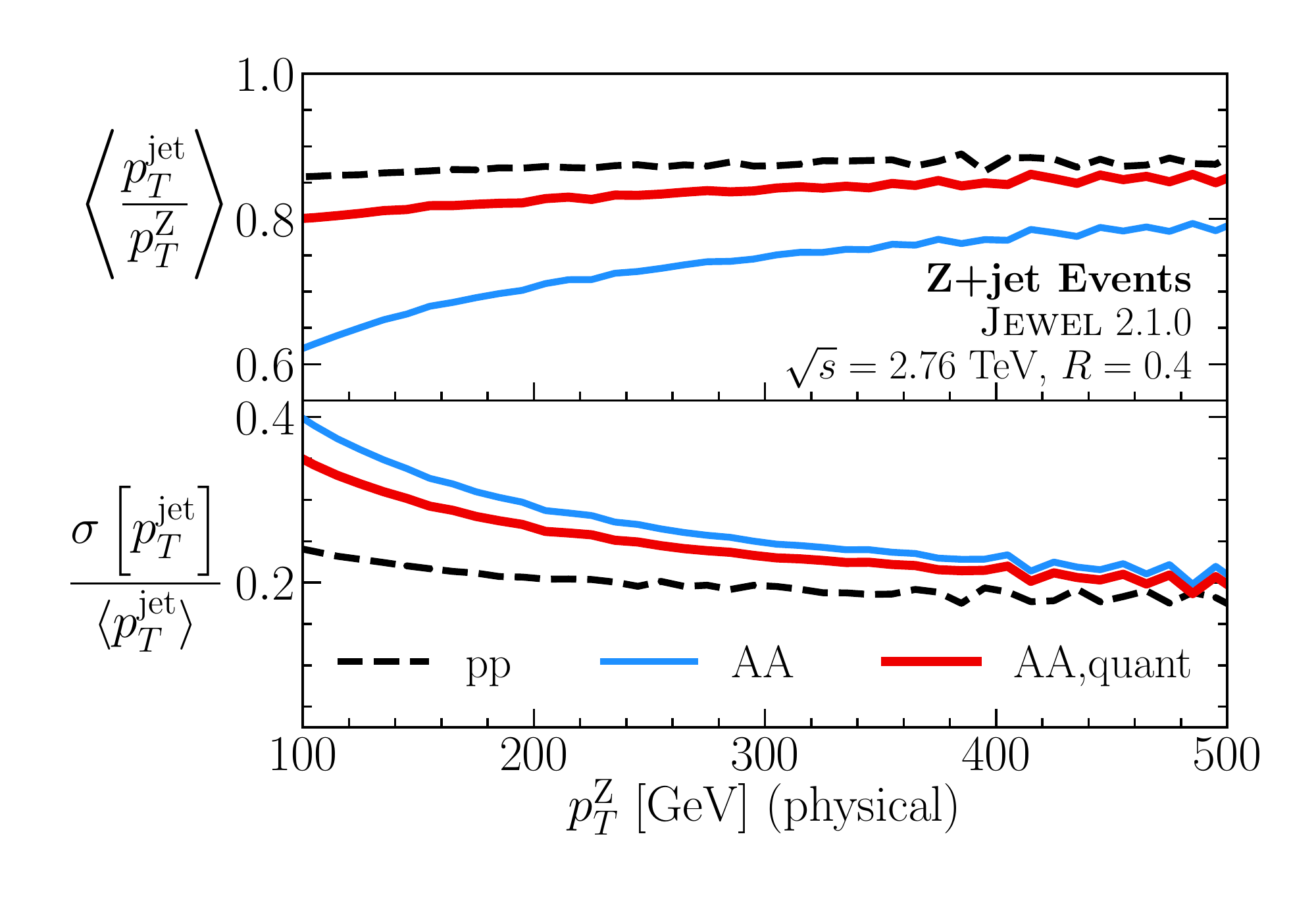}%
}
\subfloat[\label{fig:pT-vs-pTin_b}]{%
  \includegraphics[width=.5\linewidth]{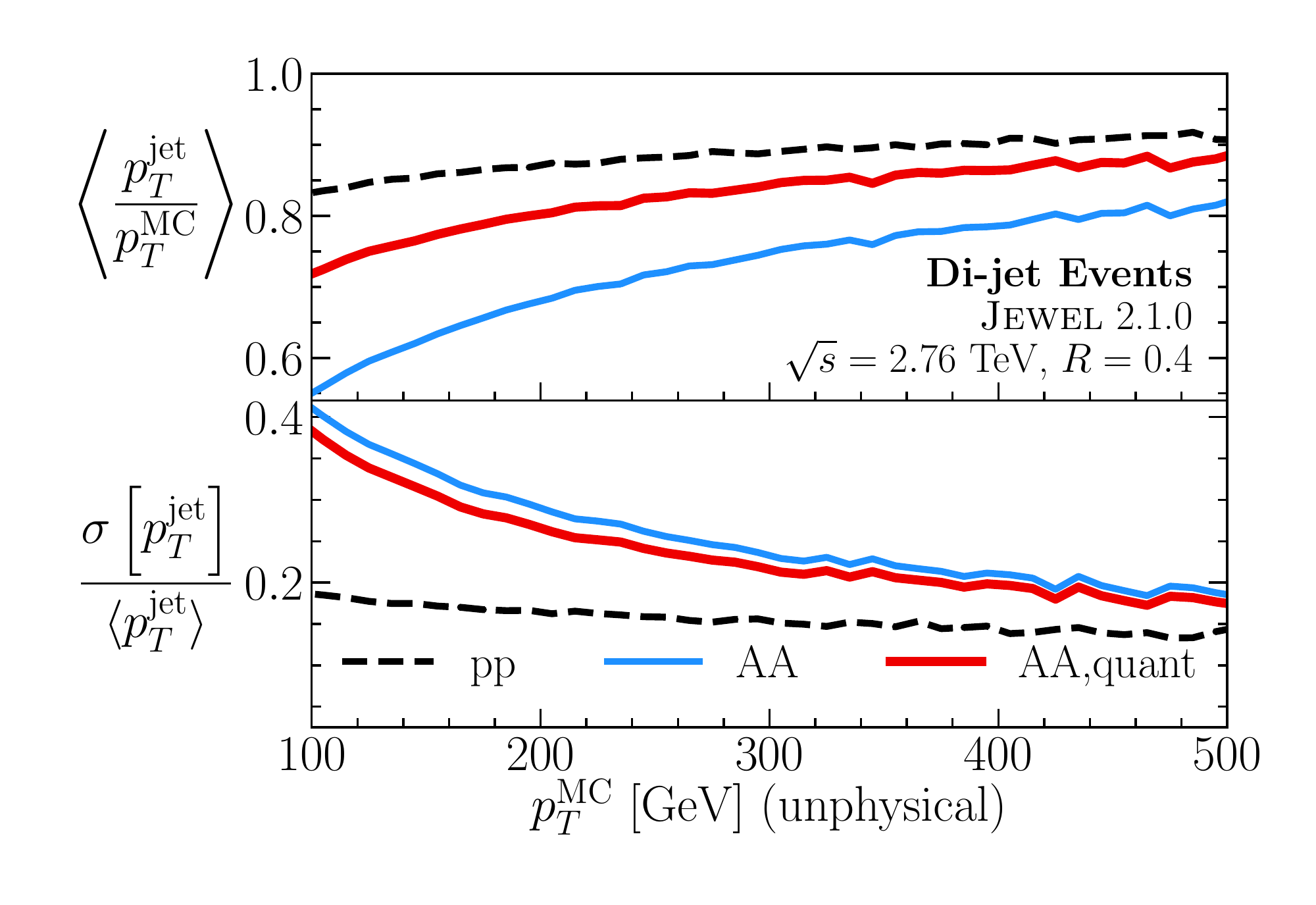}%
}
\caption{
Mean of the jet $p_T$ distribution compared to a baseline initial $p_T$ (top), along with the corresponding standard deviation (bottom).
Shown are (a) $\Z$+jet events where the baseline is the physically observable $p_T$ of the recoiling $\Z$ boson and (b) di-jet events where the baseline is the unphysical and unobservable $p^{\rm MC}_T$ of the initial hard scattering obtained from \textsc{Jewel}.
The reconstructed jet $p_T$ for proton-proton and heavy-ion jets are shown in dashed black and blue, respectively.
The $\pTq$ of the heavy-ion sample, shown in red, more closely matches the initial jet $p_T$ than the reconstructed heavy-ion $p_T$ does.}
\label{fig:pT-vs-pTin}
\end{figure*}

We now turn to validating the interpretation of $\pTq$ as a proxy for the initial $p_T$ of a heavy-ion jet before quenching by the medium.
In $\Z$+jet events, $\pTZ$ can be used as a baseline for the (approximate) initial $p_T$ of the leading recoiling jet, since the $\Z$ boson does not interact with the quark-gluon plasma.
For a given value of $\pTZ$, there is a distribution of recoil jet momenta whose mean is shown in the upper panel of \Fig{fig:pT-vs-pTin_a}.
Even in proton-proton collisions, the recoiling jet $p_T$ is systematically lower on average than $\pTZ$ due to out-of-cone radiation and events with multiple jets.
In heavy-ion collisions, it is even lower due to energy loss.
Intriguingly, the mean value of $\pTq$ (red) is much more comparable to that of $\pTvac$ (dashed black) than $\pTHI$ (blue) is, indicating that $\pTq$ is a good proxy for the initial jet $p_T$. 
On the other hand, the standard deviation of $\pTq$, shown in the lower panel of \Fig{fig:pT-vs-pTin_a}, is higher than that of $\pTvac$ due to energy loss fluctuations.
These cannot be undone by the quantile procedure, which can only give a perfect reconstruction of the distribution of $\pTvac$ in the case of strictly monotonic energy loss.

We emphasize that the distribution in \Fig{fig:pT-vs-pTin_a} is physically observable and could be used to validate the quantile procedure in experimental data.
Crucially, quantile matching can also provide a baseline for the initial jet $p_T$ in general jet events.
To validate this in di-jet events at the generator level, we use the $p_T$ of the partons from the initial hard matrix element in \textsc{Jewel}, $\pTMC$, as an (unphysical and unobservable) baseline for the initial jet $p_T$ (see \cite{Milhano:2015mng}).
We consider the two highest-$p_T$ jets and match each jet with the $\pTMC$ that minimizes $\Delta R = \sqrt{\Delta \eta^2 + \Delta \phi^2}$ between the jet and the parton.
Each of the two jets then enters independently in \Fig{fig:pT-vs-pTin_b}, which demonstrates the correlation of the jet $p_T$ to $\pTMC$ for proton-proton and heavy-ion jets, with the results of the quantile procedure in red.
\Fig{fig:pT-vs-pTin_b} is the only figure in this work that involves an unobservable quantity, and it shows remarkably similar features to \Fig{fig:pT-vs-pTin_a} which can be measured experimentally.

It might be surprising that the curves in \Fig{fig:pT-vs-pTin} are fairly flat as a function of the baseline initial $p_T$.
This can be understood, however, from a minimal model in which the final energy of a jet is obtained from its initial energy via gaussian smearing.
Consider the probability distribution
\begin{multline}
	\label{eq:pTHI-given-pTin}
	p( \pTHI | \pTin) = \int \df \pTvac \,  \mathcal{N}( \pTHI | \tilde{\mu}_2 \pTvac,\tilde{\sigma}_2 \pTvac) \\
	\times \mathcal{N}(\pTvac | \tilde{\mu}_1 \pTin,\tilde{\sigma}_1 \pTin).
\end{multline}
Here, $\mathcal{N}(x | \mu,\sigma)$ is a normal distribution in the variable $x$ with mean $\mu$ and standard deviation $\sigma$, and $\tilde{\mu}_{1,2}$ and $\tilde{\sigma}_{1,2}$ are dimensionless constants.
\Eq{eq:pTHI-given-pTin} describes the probabilistic relation between the seed-parton momentum $\pTin$ (interpreted as $\pTZ$ or $\pTMC$) and the quenched momentum $\pTHI$ via two stages of gaussian smearing: first from $\pTin$ to the unquenched jet momentum $\pTvac$, and then from $\pTvac$ to the quenched momentum $\pTHI$.
Integrating over intermediate values of $\pTvac$ gives $p(\pTHI|\pTin)$, the probability of $\pTHI$ for fixed $\pTin$.
This is an example of a model in which the average energy loss is monotonic in $p_T$, since $\mu_2 = \tilde{\mu}_2 \, \pTvac$ is a monotonic function of $\pTvac$, but energy loss is not monotonic in $p_T$ jet-by-jet since $\tilde{\sigma}_2 \neq 0$.

The mean and standard deviation of the distribution in \Eq{eq:pTHI-given-pTin} can be calculated analytically (see \cite{Bishop:2006:PRM:1162264}): 
\begin{align}
\begin{split}
	\label{eq:mean-and-std}
	\langle \pTHI/ \pTin \rangle &= \tilde{\mu}_1 \, \tilde{\mu}_2,\\
	\sigma \left( \pTHI/ \pTin \right) &= \sqrt{ \tilde{\mu}_1^2 \, \tilde{\sigma}_2^2 + \tilde{\mu}_2^2 \, \tilde{\sigma}_1^2 + \tilde{\sigma}_1^2 \, \tilde{\sigma}_2^2},
	\end{split}
\end{align}
though the resulting distribution is not generally gaussian.
These can be compared to the upper and lower panels, respectively, of \Fig{fig:pT-vs-pTin}.
The fact that \Eq{eq:mean-and-std} has no $\pTin$-dependence is consistent with the fact that the curves in \Fig{fig:pT-vs-pTin} are approximately flat.
To the extent that this model is semi-realistic, \Eq{eq:mean-and-std} and a measurement of \Fig{fig:pT-vs-pTin_a} would provide an estimate of the average energy loss and the size of energy loss fluctuations.
Taking approximate values from \Fig{fig:pT-vs-pTin_a} at $\pTZ = \SI{300}{GeV}$ of $\langle \pTvac/ \pTZ \rangle \equiv \tilde{\mu}_1 \approx 0.87$, $\sigma \left( \pTvac/\pTin \right) \equiv \tilde{\sigma}_1 \approx 0.2\,\tilde{\mu}_1 = 0.17$, $\langle \pTHI/ \pTZ \rangle \approx 0.74$, and $\sigma \left( \pTHI/ \pTZ \right) \approx 0.24 \, \langle \pTHI/ \pTZ \rangle = 0.18$, \Eq{eq:mean-and-std} yields $\tilde{\mu}_2 \approx 0.85$ and $\tilde{\sigma}_2 \approx 0.12$.
It is satisfying that this extracted $\tilde{\mu}_2$ value is comparable to $Q_{\rm AA}$ in \Fig{fig:pTloss_b}, which is a more direct proxy for fractional energy loss.

\begin{figure}[t]
\includegraphics[width=7.5cm]{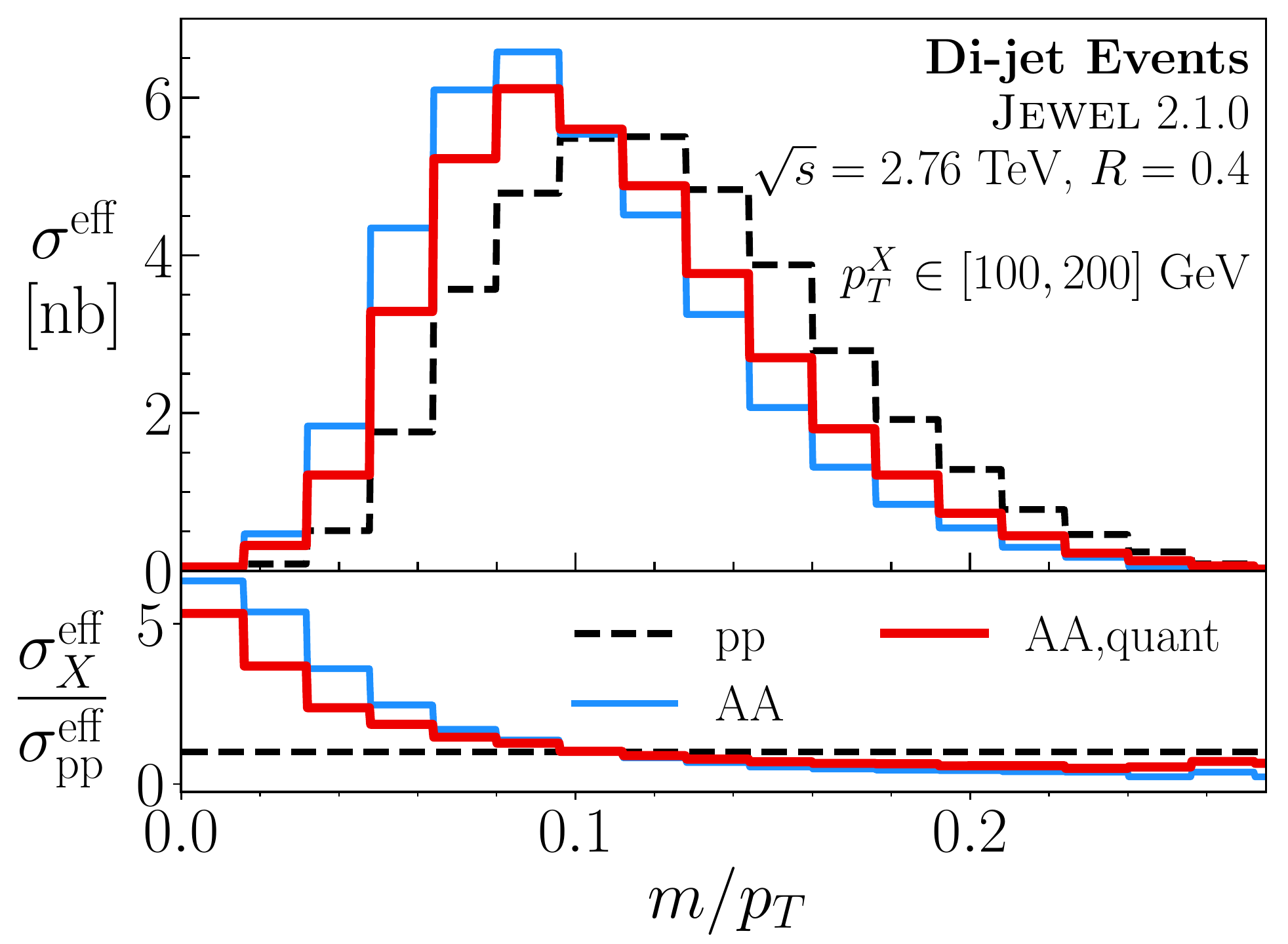}
\captionsetup{font=small,justification=raggedright}
\caption{\label{fig:moverpT} Distribution of $m/p_T$ for proton-proton (dashed black) and heavy-ion (blue) jets in di-jet events with reconstructed $p_T \in [100,200]$~GeV. Heavy-ion jets with $\pTq \in [100,200]$~GeV, corresponding to $\pTHI \in [80,173]$~GeV, are in red.
The heavy-ion result is normalized to match the proton-proton baseline but the quantile result has the correct normalization by construction.
Partially compensating for $p_T$ migration via the quantile procedure shifts $m/p_T$ towards being less modified.}
\end{figure}

As a final application in this Letter, we demonstrate how the quantile procedure can be used to characterize the effects of $p_T$ migration via an example jet substructure observable, the dimensionless ratio of the jet mass to its reconstructed $p_T$, $m/p_T$.
\Fig{fig:moverpT} shows distributions of $m/p_T$ for proton-proton and heavy-ion jets in a range of reconstructed $p_T$ in dashed black and blue, respectively.
Heavy-ion jets with that range of $\pTq$ are those in the same quantile as the proton-proton baseline, and $m/p_T$ for that sample is shown in red.
For the purpose of this example, we define $m/p_T$ from the reconstructed jet mass and $p_T$, such that the effect of the quantile procedure is only to change the $p_T$ range of jets in the selection.
Using the quantile procedure to (partially) account for the migration of jets to lower $p_T$, the red distribution shifts toward $m/p_T$ being less modified.
We note that the jet mass is known to have significant corrections from medium response \cite{KunnawalkamElayavalli:2017hxo,Park:2018acg} so this should be taken only as an illustrative example.

In conclusion, we introduced a new strategy for comparing heavy-ion jets to a baseline of proton-proton jets in the same quantile when sorted by $p_T$. 
As shown in \Fig{fig:pTloss}, our new $Q_{\rm AA}$ observable is based on the same jet $p_T$ spectra as $R_{\rm AA}$ but exposes different and complementary information. 
As shown in \Fig{fig:pT-vs-pTin}, our new $\pTq$ observable is closely correlated with the initial $p_T$ a heavy-ion jet had before energy loss to the plasma.
Thus, the quantile procedure provides a data-driven way to study the modification of quenched jets and minimize the effects of sample migration.
Experimental tests in $\Z$+jet or $\gamma$+jet can validate the effectiveness of $\pTq$ as a proxy for the initial $p_T$ of a heavy-ion jet.
If these tests are successful, the quantile procedure can then be used to re-analyze measurements of jet modification observables in general jet events with an aim toward characterizing and minimizing $p_T$ migration effects and thus compare jet samples that were born alike. 
The measurement of $Q_{\rm AA}$ will provide information on the functional form of the average energy loss which would further constrain theoretical models.
It can also be used to measure differences in average energy loss between quark- and gluon-dominated jet samples.
Measurements of $Q_{\rm AA}$ with jet grooming \cite{Butterworth:2008iy,Ellis:2009me,Krohn:2009th,Dasgupta:2013ihk,Larkoski:2014wba} may also elucidate, for example, how energy is lost by the hard core of a jet compared to the diffuse periphery.
It would also be interesting to study the application of this procedure to understanding energy loss fluctuations.
Finally, \Fig{fig:quant} shows two additional observables---the pseudo-ratio $\widetilde{R}_{\rm AA}$ and pseudo-quantile $\widetilde{Q}_{\rm AA}$---which may be relevant for experimental applications.

\begin{acknowledgments}
We thank Liliana Apolin\'ario, Yang-Ting Chien, Yen-Jie Lee, William Lewis, James Mulligan, Aditya Parikh, Krishna Rajagopal, Gavin Salam, Andrew Turner, and Korinna Zapp for helpful discussions.
JB and JT are supported by the U.S. Department of Energy, Office of Science, Office of Nuclear Physics under grant Contract Number DE-SC0011090 and the Office of High Energy Physics under grant Contract Number DE-SC0012567. 
JT is also supported by the Simons Foundation through a Simons Fellowship in Theoretical Physics, and he thanks the Harvard Center for the Fundamental Laws of Nature for hospitality while this work was completed.
GM is supported by Funda\c c\~ao para a Ci\^encia e a Tecnologia (Portugal) under project CERN/FIS-PAR/0022/2017, and he gratefully acknowledges the hospitality of the CERN theory group.
\end{acknowledgments}

\bibliographystyle{apsrev} 
\bibliography{quantile}

\end{document}